\documentclass[twocolumn]{aastex6}
\begin{document}


\title{Li in Open Clusters: Cool Dwarfs in the Young, Subsolar \\ Metallicity Cluster M35 (NGC 2168)}
\author{Barbara J. Anthony-Twarog}
\affil{Department of Physics and Astronomy, University of Kansas, Lawrence, KS 66045-7582, USA}
\email{bjat@ku.edu}
\author{Constantine P. Deliyannis}
\affil{Department of Astronomy, Indiana University, Bloomington, IN 47405-7105, USA}
\email{cdeliyan@indiana.edu}
\author{Dianne Harmer}
\affil{National Optical Astronomy Observatory, Tucson, AZ 85719, USA}
\email{diharmer@noao.edu}
\author{Donald B. Lee-Brown}
\affil{Department of Physics and Astronomy, University of Kansas, Lawrence, KS 66045-7582, USA}
\email{dbleebrown@gmail.com}
\author{Aaron Steinhauer}
\affil{ Department of Physics \& Astronomy, SUNY Geneseo, 1 College Circle, Geneseo, NY 14454, USA}
\email{steinhau@geneseo.edu}
\author{Qinghui Sun}
\affil{Department of Astronomy, Indiana University, Bloomington, IN 47405-7105, USA}
\email{qingsun@indiana.edu}
\and
\author{Bruce A. Twarog}
\affil{Department of Physics and Astronomy, University of Kansas, Lawrence, KS 66045-7582, USA}
\email{btwarog@ku.edu}


\begin{abstract}
Hydra spectra of 85 G-K dwarfs in the young cluster, M35, near the Li 6708 \AA\ line region are analyzed. From velocities and {\it Gaia} astrometry, 78 are likely single-star members which, combined with previous work, produces 108 members with $T_{\mathrm{eff}}$ ranging from 6150 to 4000 K as defined by multicolor, broad-band photometry, $E(B-V)$ = 0.20 and [Fe/H] = -0.15, though there are indications the metallicity may be closer to solar. A(Li) follows a well-delineated decline from 3.15 for the hottest stars to upper limits 
$\leq 1.0$ among the coolest dwarfs. Contrary to earlier work, M35 includes single stars at systematically higher A(Li) than the mean cluster relation. This subset exhibits higher $V_{ROT}$ than the more Li-depleted sample and, from photometric rotation periods, is dominated by stars classed as {\it convective} (C); all others are {\it interface} (I) stars. The cool, high-Li rapid rotators are consistent with models that consider simultaneously rapid rotation and radius inflation; rapid rotators hotter than the sun exhibit excess Li depletion, as predicted by the models. The A(Li) distribution with color and rotation period, when compared to the Hyades/Praesepe and the Pleiades, is consistent with gyrochronological analysis placing M35's age between the older M34 and younger Pleiades. However, the Pleiades display a more excessive range in A(Li) and rotation period than M35 on the low-Li, slow-rotation side of the distribution, with supposedly younger stars at a given $T_{\mathrm{eff}}$ in the Pleiades spinning slower, with A(Li) reduced by more than a factor of four compared to M35.

\end{abstract}





\section{Introduction}

Li evolution as a process has dual connotations: the time-dependent variation in the abundance of Li within the observed atmosphere of a star as it goes from its pre-main sequence origins to stellar death among the giants, and the alteration over time, place, and metallicity of the primordial abundance of the interstellar medium from which a star may form. They are clearly coupled because the latter supplies the initial conditions which help constrain the former. Since the timescale for formation is rapid compared to the lifetime of a star, it can be a challenge to observe a star before its evolution alters the primordial value, unless one knows that the impact of time on the atmospheric Li abundance of a star of a given mass should be small to nonexistent. Standard stellar evolution theory (SSET) \citep{PI97} predicts that the primary mechanism for apparent Li-depletion, the nuclear destruction of Li at temperatures above $\sim 2.5 \times 10^6$ K, should be irrelevant for main sequence stars in the $T_{\mathrm{eff}}$ regime above $\sim$6000 K due to the predicted lack of significant mixing/convection within the atmospheres of these stars. The existence of the well-studied Li-dip near $T_{\mathrm{eff}}$ $\sim$ 6600 K \citep[see, e.g.][among many others]{BT86, CH01, PA04, AT09, RA12, CU12, CU17}  has effectively destroyed any hope that the SSET models at these temperatures supply realistic assessments of their true atmospheric nature, though observations of field and cluster stars above $T_{\mathrm{eff}}$ $\sim$ 6800 K give some hope that the assumption might still hold. In this hotter regime, however, the absence of atmospheric instability points toward a second mechanism for altering the atmospheric abundance, diffusion. This process is regularly proposed to explain the perceived discrepancy between the predicted cosmological A(Li) (defined as $12 +$ log [Li/H])
from observations of the cosmic background radiation \citep{CO14} and the systematically lower observed value in turnoff and subgiant stars of metal-poor globular clusters \citep{NO12, GR13, GR14, GR16}. At minimum, there is no disagreement that the Li abundance of the interstellar medium has changed by approximately an order of magnitude since the formation of the Milky Way halo given the observed solar system value \citep{GR98} and that of clusters like the Hyades and Praesepe \citep{CU17}; the temporal, spatial, and metallicity dependences, if any, of this change remain unsettled.

This litany of uncertainty is presented to emphasize why cluster observations have taken on such a critical role in disentangling the impact of the multitude of factors which can influence the Li abundance of a given star. While age and mass estimates can be derived for individual hotter, higher mass stars with known metallicity and reddening that undergo substantial evolution on timescales shorter than the age of the Universe, for cooler dwarfs one is often left with inferred ages based upon secondary parameters such as rotation rate \citep[e.g.][]{RE15, ME15} or A(Li) itself \citep[e.g.][]{ST03b, CU11, KR14}. The often-stated power of cluster samples is that the age, metallicity, reddening, and distance for the cooler stars can be extracted from the observationally more pliant sample of hotter stars within the same cluster. A critical weakness in this approach is that the range of clusters near enough to be observed spectroscopically at the cooler end of the main sequence is limited in metallicity, especially at the younger end of the scale where relatively unevolved stars still on the main sequence both above (brighter than) and below (fainter than) the Li-dip can be evaluated within the same object. Examples of such studies can be seen in \citet{ST03b}, \citet{SE03}, \citet{CU11}, and \citet{CU17}.

The focus of the current investigation within the WIYN Open Cluster Survey\footnote{This paper is number 77 in the WOCS series.} (WOCS; \citet{MA00}) is M35 (NGC 2168), a rich, northern open cluster with an age well below that of the Hyades and, most important, a metallicity which places it approximately a factor of two below the Hyades. The specific motivations for this study come out of the differences between the Li-$T_{\mathrm{eff}}$-rotation patterns observed in M35 by \citet{BA01} and those observed previously in the Pleiades.  In particular, proportionately fewer rapidly rotating G and K dwarfs were found in M35, and perhaps not coincidentally, almost no high-Li G and K dwarfs above the mean Li-$T_{\mathrm{eff}}$ trend defined by slower rotators.  Are these real trends that require explanation, or are they simply artifacts of a small sample size?  The present study was undertaken as an attempt to address these questions by expanding the sample size. The age-metallicity combination supplied secondary motivation for reaching well beyond the solar neighborhood and closer to the fainter observational limits of the Hydra spectrograph to probe the sensitivity of Li evolution to metallicity among cooler dwarfs. The cluster has recently taken on added significance \citep{LI16,So17} because of its inclusion within the {\it Kepler} K2 mission \citep{HO14}.

The outline of the paper is as follows: Section 2 discusses the Hydra observations of the sample selected for analysis within M35; Section 3 lays out the current state of our knowledge of the basic cluster parameters for M35: metallicity, reddening, age, and distance, using recent input from {\it Gaia} Data Release 2 (DR2) \citep{GA16, GA18b} to aid in isolating cluster members from field stars. Section 4 contains the derivation of the Li abundances and their implication for M35, for Li evolution relative to other clusters, and for current models probing the relation between A(Li) and rotation. Section 5 is a summary of our conclusions.

\section{Observations and Data Reduction}

\subsection{Sample Selection}
Two previous Hydra spectroscopic studies of Li in M35 have been published. \citet{BA01} emphasized 76 cooler dwarfs between $V$ = 17.7 and 14.6 while \citet{ST03b} and \citet{SD04} concentrated on 165 hotter dwarfs between $V$ = 12.25 and 15, mapping primarily the region of the Li-dip. The overlap between these two studies consists of only two stars. Moreover, of the 76 cool dwarfs selected by the earlier study, radial velocities \citep{BA01, GE10, LE15} eliminate all but 38 from the sample due to nonmembership and/or binarity. One star, 5190 on the system of \citet{BA01b}, has not been identified in any of the proper-motion, radial-velocity or alternative photometric studies, despite the relatively bright $V$ of 15.7. While \citet{BA01} find a radial velocity consistent with cluster membership, their derived Li abundance is anomalously low for a cluster star of its color, implying that it is almost certainly an older field star. This star will be excluded from further discussion, reducing the preliminary \citet{BA01} sample to 37 single members. The 50\% success rate for membership isn't unexpected given that the first sample was constructed solely from the $VRI$ photometry of \citet{BA01b} since the sample was fainter than the faint limit of the \citet{MC86} proper-motion study. 

For the current study, the availability of some preliminary radial-velocity measures as part of the WOCS sample (R. Mathieu 2000, private communication) that eventually produced \citet{GE10} and \citet{LE15} greatly enhanced the success rate for membership, especially in conjunction with the newer, precision $UBVRI$ photometric survey under way at the time \citep[][in preparation]{DE18} which allowed selection of stars close to the single-star fiducial sequence from $UBVRI$ data to help minimize contamination by nonmembers and binaries. Of the 85 stars in our sample, 77 stars brighter than $V$ = 16.7 had preliminary radial velocities consistent with membership; of the 74 included in the extensive survey by \citet{LE15}, 73 are now confirmed to be single, radial-velocity members with only one tagged as a binary member. The additional 8 stars were selected between $V$ = 16.7 and 17.9 based upon their location in the CMD. As we will discuss below, 6 of these 8 have radial velocities from our current observations consistent with single-star cluster membership, one does not, and the spectrum for the faintest star generated an unreliable velocity estimate. The 85 stars which comprise the current sample are listed in Table 1. For 74 stars, the identification number listed is the WOCS ID; the remaining 11 without WOCS ID's have been assigned the designation F1 through F11. The coordinates for all the stars are listed in columns 2 and 3 and were used to cross-reference the current sample with that of \citet{LE15} to obtain appropriate WOCS ID's.

\floattable
\begin{deluxetable}{ccccchcchcchcchcchccccc}
\tablenum{1}
\tablecaption{Photometry for M35 Stars}
\tablewidth{0pt}
\tabletypesize\footnotesize
\tablewidth{0pc}
\tablehead{
\colhead{WOCS ID} & \colhead{$\alpha(2000)$} & \colhead{$\delta(2000)$} & 
\colhead{$\rm{U}$} & \colhead{$\sigma_{U}$} & \nocolhead{$\sigma_{2U}$}  &  
\colhead{$\rm{B}$} & \colhead{$\sigma_{B}$} & \nocolhead{$\sigma_{2B}$}  &  
\colhead{$\rm{V}$} & \colhead{$\sigma_{V}$} & \nocolhead{$\sigma_{2V}$}  &  
\colhead{$\rm{R}$} & \colhead{$\sigma_{R}$} & \nocolhead{$\sigma_{2R}$}  &  
\colhead{$\rm{I}$} & \colhead{$\sigma_{I}$} & \nocolhead{$\sigma_{2I}$}  &  
\colhead{$N_U$} & \colhead{$N_B$} & \colhead{$N_V$} & \colhead{$N_R$} & \colhead{$N_I$} }
\startdata
 16010 & 92.22063 & 24.28908 & 15.375 & 0.008 & 0.007 & 15.256 & 0.008 & 0.007 & 14.514 & 0.007 & 0.012 & 14.049 & 0.008 & 0.010 & 13.586 & 0.008 & 0.019 & 9 & 12 & 11 & 10 & 10 \\
 32048 & 91.85886 & 24.44003 & .... & .... & .... & 15.366 & 0.015 & 0.005 & 14.590 & 0.013 & 0.006 & 14.125 & 0.015 & 0.000 & 13.652 & 0.014 & 0.015 & 0 & 2 & 3 & 2 & 3 \\
 18011 & 92.37294 & 24.35100 & 15.557 & 0.008 & 0.010 & 15.427 & 0.008 & 0.010 & 14.654 & 0.007 & 0.008 & 14.192 & 0.009 & 0.004 & 13.728 & 0.008 & 0.013 & 9 & 10 & 11 & 9 & 9 \\
 22030 & 92.10622 & 24.15492 & 15.546 & 0.029 & 0.011 & 15.434 & 0.008 & 0.004 & 14.685 & 0.007 & 0.005 & 14.228 & 0.011 & 0.007 & 13.772 & 0.009 & 0.011 & 3 & 8 & 10 & 7 & 7 \\
 31038 & 92.10316 & 24.07828 & .... & .... & .... & 15.473 & 0.019 & 0.006 & 14.763 & 0.016 & 0.005 & 14.312 & 0.021 & 0.021 & 13.891 & 0.020 & 0.012 & 0 & 2 & 3 & 2 & 3 \\
 13006 & 92.27103 & 24.29774 & 15.736 & 0.009 & 0.010 & 15.552 & 0.008 & 0.012 & 14.778 & 0.007 & 0.025 & 14.273 & 0.009 & 0.016 & 13.798 & 0.008 & 0.015 & 7 & 11 & 10 & 9 & 9 \\
 15007 & 92.33062 & 24.31891 & 15.752 & 0.008 & 0.009 & 15.594 & 0.007 & 0.013 & 14.783 & 0.006 & 0.012 & 14.287 & 0.008 & 0.011 & 13.809 & 0.008 & 0.013 & 8 & 13 & 13 & 11 & 11 \\
 36048 & 91.86958 & 24.22800 & .... & .... & .... & 15.526 & 0.018 & 0.004 & 14.794 & 0.015 & 0.011 & 14.335 & 0.018 & 0.013 & 13.880 & 0.022 & 0.012 & 0 & 2 & 3 & 3 & 3 \\
 31026 & 92.08678 & 24.45520 & 15.654 & 0.009 & 0.014 & 15.553 & 0.010 & 0.006 & 14.808 & 0.010 & 0.008 & 14.349 & 0.009 & 0.008 & 13.918 & 0.010 & 0.031 & 8 & 7 & 8 & 10 & 10 \\
 33034 & 92.21014 & 24.61621 & .... & .... & .... & 15.551 & 0.017 & 0.054 & 14.810 & 0.013 & 0.021 & 14.359 & 0.015 & 0.018 & 13.889 & 0.013 & 0.016 & 0 & 3 & 6 & 6 & 6 \\
\enddata
\tablecomments{Sample lines for Table 1 are presented here to illustrate the content of the table, which will be available in full online.}
\end{deluxetable}

\subsection{Hydra Observations}
Spectroscopic data were obtained using the WIYN 3.5m telescope\footnote{The WIYN Observatory was a joint facility of the University of Wisconsin-Madison, Indiana University, Yale University, and the National Optical Astronomy Observatory.} and the Hydra multi-object spectrograph over 3 nights in March 2001. The same configuration of 85 star fibers with 8 sky fibers was repeated twice each night for 2 hours each, with a cumulative total of 12 hours for the configuration. Four to five bright radial-velocity standards were observed each night and a complete set of solar/sky observations through all functional fibers was obtained each afternoon. Comparison Thorium-Argon lamp spectra were recorded periodically throughout each night for wavelength calibration purposes. Our previous studies employed the echelle 316 l/mm grating, which provides R $\sim$ 13,000 with the blue fibers and R $\sim$ 17,000 with the red fibers near the Li I 6707.8 \AA\ wavelength region, covering a range of $\sim$400 \AA.  For this study, we employed the 31.6 l/mm KPNO coude grating in order 84, hoping to take advantage of a higher throughput efficiency and slightly higher resolution. However, due to the change in configuration, the usable spectra within M35 only cover the 6700 \AA\  to 6730 \AA\  range. This limitation has no impact on the measurement of Li, but severely reduces the prospective number of metal lines available for measurement to constrain the metallicity and makes use of our recently developed neural network approach for abundances, ANNA \citep[][in preparation]{AT18, LB18}, unreliable. Despite the narrow bandpass, our spectra have a resolution of $R \sim$ 19000, with 166 m\AA/pixels. Details on the processing and reduction procedures from raw spectra to normalized, wavelength-calibrated spectra can be found in a number of our previous investigations \citep{AT09, CU12, LB15} and will not be repeated here. 

\subsection{Radial Velocities and Rotation}

Individual heliocentric stellar radial velocities, $V_{RAD}$, were derived from each summed composite spectrum utilizing the Fourier-transform, cross-correlation facility {\it fxcor} in IRAF\footnote{IRAF is distributed by the National Optical Astronomy Observatory, which is operated by the Association of Universities for Research in Astronomy, Inc., under cooperative agreement with the National Science Foundation.} \citep{DT86,DT93}. In this utility, program stars are compared to stellar templates of similar effective temperature ($T_{\mathrm{eff}}$) over the full wavelength range of our spectra. Output of the {\it fxcor} utility characterizes the cross-correlation function, from which estimates of each star's radial velocity are inferred. Rotational velocities can also be estimated from the cross correlation function full-width (CCF FWHM) using a procedure developed by \citet{ST03b}. This procedure exploits the relationship between the CCF FWHM, line widths and $V_{ROT}$, using a set of numerically ``spun up" standard spectra with comparable spectral types to constrain the relationship. For simplicity, $V_{ROT}$ as used here implicitly includes the unknown $sin$ $i$ term. As evaluated below, the precision of our final $V_{RAD}$ estimates for each star exhibits minor impact due to the narrower-than-usual bandpass of our Hydra spectra, but the reliability of the $V_{ROT}$ estimates declines measurably and the stars with the largest $V_{ROT}$ are also much more likely to generate implausible $V_{RAD}$. Table 2 lists the $V_{RAD}$ and $V_{ROT}$ results for our sample from \citet{LE15} and from our measured spectra. Proper-motion and radial-velocity membership probabilities from previous work are also supplied.

Of the 85 stars in our sample, 74 overlap with the extensive radial-velocity survey of \citet{LE15}. Of these, one is classified as a binary member while for three others, {\it fxcor} generated clearly implausible and discrepant values from our spectra. Of these three, two are classed as rapid rotators by \citet{LE15}, which may explain in part the failure to find a plausible velocity. From the 71 remaining stars, the mean residual in velocity, in the sense (LE - Table 2), is -0.22 $\pm$ 0.15 km/sec ($\sigma_{\mu}$), only slightly larger than the error predicted from the individual precision of the radial velocity measurements, 0.13 km/sec. The mean radial velocity of the cluster from the 73 single stars of \citet{LE15} is -8.07 $\pm$ 0.10 km/sec ($\sigma_{\mu}$). Excluding the known binary and one star fainter than the limit of the \citet{LE15} survey, the 78 stars in the current investigation with measurable radial velocities have a mean of -7.88 $\pm$ 0.17 km/sec ($\sigma_{\mu}$). While radial velocities aren't available from {\it Gaia} DR2 for any stars within the current investigation, for the brighter cluster stars, \citet{GA18b} quote -7.70 $\pm$ 0.26 km/sec for the cluster. It is encouraging to note that the one star which deviates the most from the cluster mean in both investigations is the binary, WOCS 14801. One star, F8, at $V$ = 17.56 has a derived radial velocity of -18.49 km/sec. Whether this star is a binary, nonmember, or simply a measurement error at faint magnitudes remains unknown. For purposes of our discussion, it will be classed as a radial-velocity nonmember/binary. One star (F3) did have a preliminary radial velocity as part of the database used to construct the original sample but the spectrum from the current investigation produced an implausible {\it fxcor} value and it is too faint for the \citet{LE15} database. We have included the original $V_{RAD}$ in Table 2 with an appropriate indication of its uncertainty.

\floattable
\begin{deluxetable}{rrrrrrrrrrrrrrrr}
\tablenum{2}
\tablecaption{Basic Information for M35 Stars}
\tablewidth{0pt}
\tabletypesize\small
\tablewidth{0pc}
\tablehead{
\colhead{WOCS ID} & \colhead{$\rm ID_A$} & \colhead{$\rm ID_B$} & \colhead{$\rm ID_C$} & 
\colhead{$\%_{RV}$} & \colhead{$\rm {\%_{B}}$} & \colhead{$\%_{C}$} & \colhead{$\rm {MC}$} &
\colhead{$\rm RV_{LE}$} & \colhead{$\sigma_{RV}$} & \colhead{$\rm V_{RotLE}$} & \colhead{$\sigma{V_{RotLE}}$} &
\colhead{$\rm RV$} & \colhead{$\sigma{RV}$} &
\colhead{$\rm V_{ROT}$} & \colhead{$\sigma{V_{ROT}}$}}
\startdata 
 16010 & ... & 192 & ... & 95 & 90 & ... & SM & -8.89 & 0.39 & 11.2 & 0.4 & -8.60 & 1.05 & 17.9 & 1.3 \\
 32048 & ... & ... & ... & 95 & ... & ... & SM & -7.50 & 0.06 & 10.7 & 0.2 & -5.90 & 0.75 & 12.9 & 0.4 \\
 18011 & ... & 413 & ... & 86 & 88 & ... & SM & -6.58 & 0.68 & 15.3 & 1.2 & -5.73 & 1.11 & 13.7 & 0.8 \\
 22030 & ... & ... & 196 & 95 & ... & 67 & SM & -7.47 & 0.17 & 10.0 & 0.0 & -7.74 & 1.03 & 11.8 & 0.5 \\
 31038 & 196 & ... & ... & 96 & ... & ... & SM & -8.55 & 0.54 & 15.0 & 0.3 & -8.22 & 0.95 & 10.6 & 0.3 \\
 13006 & ... & ... & ... & 96 & ... & ... & SM & -8.29 & 0.17 & 10.0 & 0.0 & -8.42 & 0.38 & 12.8 & 0.3 \\
 15007 & ... & 355 & ... & 96 & 81 & ... & SM & -8.06 & 0.45 & 12.0 & 0.7 & -7.72 & 0.95 & 13.1 & 0.6 \\
 36048 & ... & ... & ... & 85 & ... & ... & SM & -6.52 & 0.67 & 10.3 & 0.3 & -6.79 & 0.70 & 10.5 & 0.5 \\
 31026 & ... & ... & 168 & 95 & ... & 90 & SM & -8.81 & 0.10 & 10.0 & 0.0 & -7.29 & 0.89 & 10.4 & 0.3 \\
 33034 & ... & ... & ... & 70 & ... & ... & SM & -10.17 & 0.40 & 11.4 & 0.7 & -8.33 & 0.88 & 12.1 & 0.4 \\
\enddata
\tablecomments{Sources for identifications and proper motion memberships corresponding to labels $\rm ID_A$, $\rm ID_B$ and $\rm ID_C$ refer respectively to \citet{ME09}, \citet{MC86} and \citet{CU71}. The source for radial-velocity membership probabilities is \citet{LE15}. Sample lines for Table 2 are presented here to illustrate the content of the table, which will be available in full online.}
\end{deluxetable}


\subsection{Proper Motion and Parallax}
As mentioned earlier, because of its greater distance than the standard nearby young open clusters, Hyades, Praesepe, and Pleiades, astrometric studies
of M35 have, until now, been limited to the brighter cluster members, thereby excluding the majority of stars within the spectroscopic survey. With the 
availability of the {\it Gaia} DR2 \citep{GA18a}, this deficiency can be remedied. We have cross-matched the previously identified cluster probable radial-velocity, single-star members with the DR2 database, identifying 36 of 37 stars from \citet{BA01} and 84 of 85 stars from Table 1. The two stars without astrometric info are BS 5407 \citep{BA01} and, ironically, star F8 from Table 1. Figure 1 shows the vector-point diagram for the two spectroscopic 
data sets,  with the sample of \citet{BA01} in blue and Table 1 in red. The cluster concentration is obvious but clearly a few stars fall away from the well-defined motion of the cluster, defined from the much larger DR2 cluster database \citep{GA18b} as $\mu$$_{RA}$ = 2.1819 $\pm$ 0.0079 (sem) mas/yr and $\mu$$_{DEC}$ = -2.9657 $\pm$ 0.0075 (sem) mas/yr. The black circle outlines a radius of 1.70 mas/yr; any star outside this zone is classed as a proper-motion nonmember. For \citet{BA01} this eliminates 7 stars (5267, 5339, 5376, 5388, 5408, 5410, and 5419), while from Table 1, only 3 stars are removed (32029, 36012, and 56021). A second constraint was imposed using the distribution of parallaxes; 5 of the 7 proper-motion nonmembers \citep{BA01} were also tagged via parallax with parallax values different from the cluster mean by more than 0.2666 mas, but no additional interlopers were identified. For Table 1, one star with a negative parallax, 86043, was eliminated. The final sample consists of 30 stars from \citet{BA01} and 79 stars from Table 1. A summary membership class, based upon proper motion, radial velocity, and parallax is given in Table 2.

\begin{figure}
\figurenum{1}
\plotone{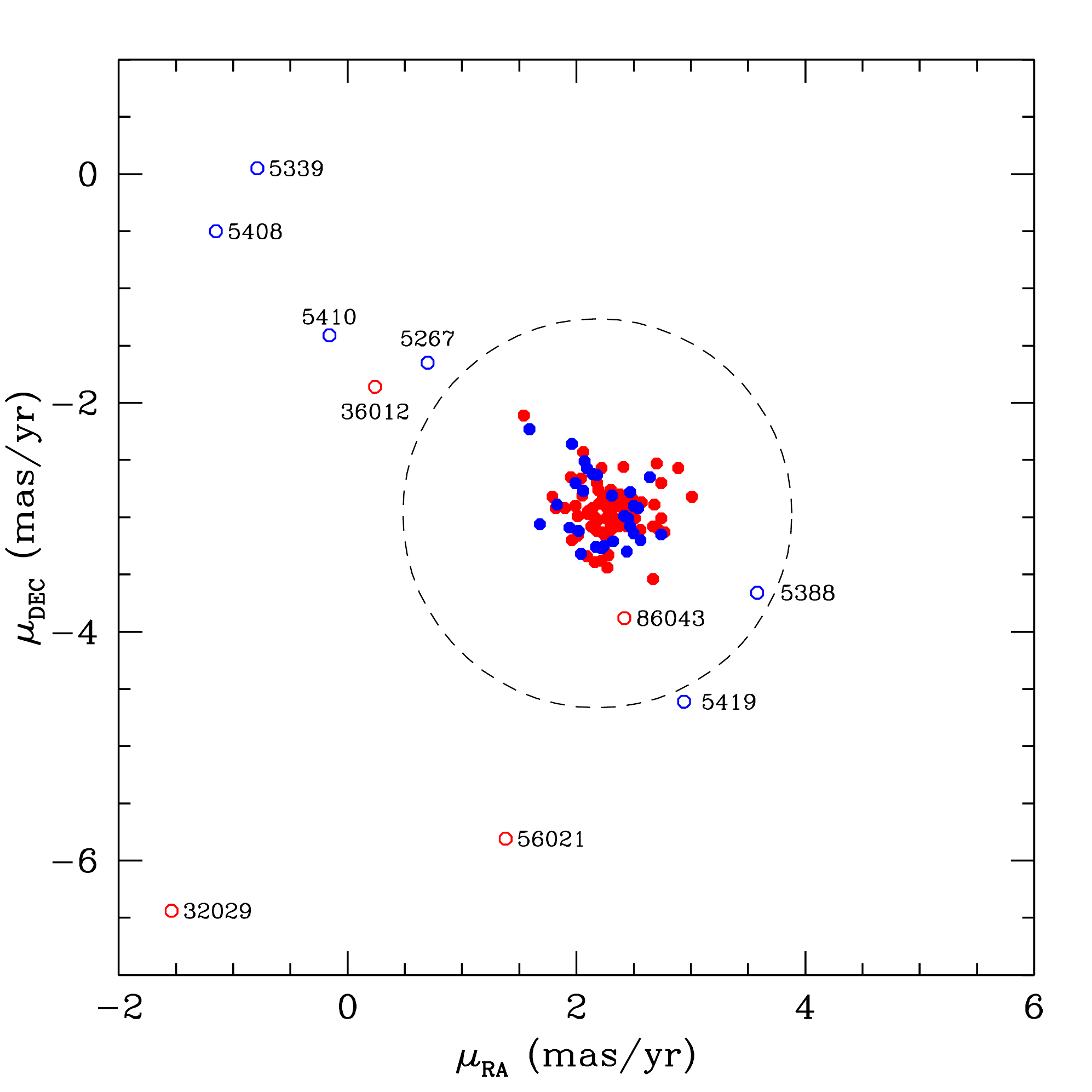}
\caption{Proper-motion vector-point diagram for radial-velocity single-star members of M35 from \citet{BA01} (blue symbols) and Table 1 (red symbols). The black
circle outlines the adopted boundary for stars classed as members. Nonmembers are tagged with their individual identification numbers.}
\end{figure}  

\subsection{UBVRI Photometry}
To get a handle on the $T_{\mathrm{eff}}$ for abundance estimation, we include in Table 1 the complete set of UBVRI CCD observations from the survey by \citet[][in preparation]{DE18}. The photometry is based upon frames taken with the former T2KA 2K $\times$ 2K CCD chip at the f/7.5 focus of the KPNO 0.9 m telescope. Each field has a 23$\arcmin$ $\times$ 23$\arcmin$ field of view, with the frames arranged in a mosaic of five fields, i.e. one central field and four tiled around the center for a total field of view of 40$\arcmin$ $\times$ 40$\arcmin$. The photometry is calibrated to the standard system as defined by \citet{LA92}.

\section{M35 - Fundamental Properties}

Categorizing the population properties of M35 stars has been the focus of extensive investigation for decades, including proper motions \citep{EB42, CU71, MC86, KH97, MC11, BO15}, radial velocities \citep{ME08, GE10, LE15}, photoelectric and CCD photometry \citep{HO61, SU92, SB99, BA01, VH02, KA03, MN07, BO15}, binary star identification \citep{MM05, ME06, ME09, LE15} and spectroscopic abundance estimation \citep{BA01, SD04}, among a long list of papers too numerous to mention. A comprehensive summary of the key cluster parameters of reddening, distance, and age as of 2003 is supplied in Table 1 of \citet{KA03}. Surprisingly, very little has changed in the intervening 15 years. Looking at analyses from the CCD era, the published reddening values range from $E(B-V)$ just under 0.20 to 0.30, while the ages span a factor of two, from 100 Myr to 200 Myr. The true distance moduli range from $(m-M)_o$ = 9.3 to 9.8. As expected, the youngest age and smallest distance are coupled to the largest adopted reddening of $E(B-V) = 0.30$. Perhaps the most surprising is the minimal information regarding the cluster metallicity. The high-dispersion spectroscopic work to date includes that of \citet{BA01}, who used 17 Fe I lines in Hydra spectra from 9 main sequence stars to obtain [Fe/H] = -0.21 $\pm$ 0.011 ($\sigma_{\mu}$), and \citet{ST03b}, who analyzed 138 Fe lines of 23 hotter dwarfs to find [Fe/H] = -0.143 $\pm$ 0.014 ($\sigma_{\mu}$). \citet{SB99} compared broad-band UBV photometry to model-based relations and $E(B-V)$ = 0.255 to conclude that [Fe/H] is $\sim$ -0.3, with a significant uncertainty. More recent UBVRI analysis by \citet[][in preparation]{DE18} using the full set of photometry which provides the color indices adopted in this investigation leads to [Fe/H] = -0.18 $\pm$ 0.05, with a reddening of $E(B-V)$ = 0.20 $\pm$ 0.01. The only additional photometric abundance of relevance comes from DDO photometry of one giant; \citet{TW97} find $E(B-V)$ = 0.19 and [Fe/H] = -0.16 $\pm$ 0.09, on a metallicity scale where NGC 752 is [Fe/H] = -0.09. As we will discuss below, there is reason to believe that M35 actually has [Fe/H] closer to solar. We will, however, adopt $E(B-V)$ = 0.20 as the cluster reddening in all future discussions. 

\subsection{Color-Magnitude Diagram and $T_{\mathrm{eff}}$}
As a basic check on the membership and single-star nature of the spectroscopic sample, we can make use of the cluster CMD to identify stars with potentially anomalous locations compared to an unevolved main sequence star within the color range of interest. To minimize the photometric errors and eventually obtain a reliable $T_{\mathrm{eff}}$ for application within the spectroscopic analysis, it was decided that use would be made of all the available photometry, rather than just selecting $(B-V)$ as the representative color. Following a pattern outlined in \citet{SD04}, six possible color indices were constructed from all the combinations of the $BVRI$ dataset for each star; $U$ wasn't included because it was unavailable for a majority of the stars. To expand the database for defining the relationships among the main sequence stars, all radial-velocity, single-star members \citep{LE15} were cross matched with DR2, and astrometric nonmembers based upon proper motion (as in Fig. 1) and/or parallax were removed. Using the combined sample from Table 1 and the astrometrically restricted \citet{LE15} data, cubic relations were derived to transform each of the 5 indices to $(B-V)$. These transformations were applied to each index for all stars in the aforementioned samples, as well as the 30 single-star members from the sample of \citet{BA01}. The pseudo $(B-V)$ indices for each star were averaged with a weighting based upon the quality of the fit between the indices. The weights ranged from 1.0 and 0.99 for $(B-V)$ and $(B-R)$, respectively, to 0.85 for $(R-I)$. The residuals for the transformed colors relative to the original $(B-V)$ produced dispersions of 0.016, 0.019, 0.042, 0.039 and 0.055 mags for $(B-R)$, $(B-I)$, $(V-R)$, $(V-I)$ and $(R-I)$, respectively, with the scatter dominated by the stars at the faintest and reddest portion of the sample distribution.

Fig. 2 illustrates the CMD from all three samples; crosses are the single-star, astrometric members with $UBVRI$ photometry culled from the \citet{LE15} set, blue open squares are the astrometric and radial-velocity members from \citet{BA01}, and the open red circles are the astrometric and radial-velocity members from the current investigation. Despite the classification as single stars based upon radial-velocity variability, it is apparent from the spread in $V$ at a fixed $(B-V)$ among the crosses that a non-negligible fraction of these stars (crosses) must have binary companions. The contrast with the open circles and squares is striking; these stars populate a tighter band in $(B-V)$, distributed to the bluer side of the main sequence, confirming that most are, in fact, highly probable single-star members, or at least binaries with a relatively low mass ratio. 

\begin{figure}
\figurenum{2}
\includegraphics[angle=270,width=\columnwidth]{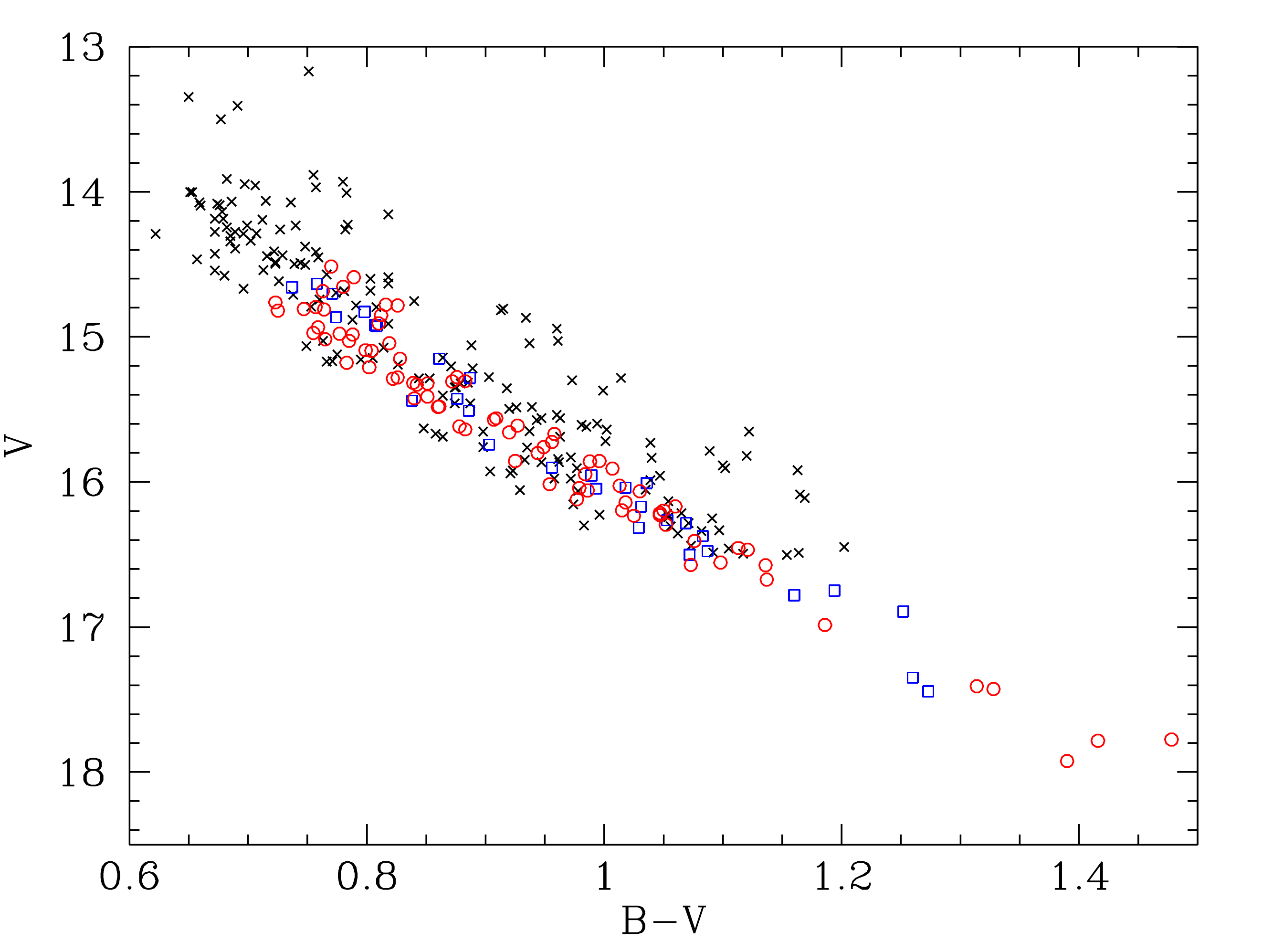}
\caption{CMD for probable members within M35. Crosses are astrometric members classed as single-star members by \citet{LE15}, open red circles are all single members in Table 2, and open blue squares are all probable single-star members from \citet{BA01}.}
\end{figure}

Fig. 3 shows the CMD of the two samples observed for Li analysis. Symbols have the same meaning as in Fig. 2 with the following exceptions. The filled cyan circle is the star in the current study classed as a binary member. The filled green symbol shows the faint star, F8, with a radial velocity inconsistent with single-star membership. The filled blue circle is the faintest star, F11, which does not have a reliable radial velocity but appears to be an astrometric member. Superposed is an isochrone of age 200 Myr from the Victoria-Regina (VR; \citet{VA06}) compilation. The isochrone has [Fe/H] = -0.14; we have incremented the VR $(B-V)$ colors by +0.01, in conformity with our past usage of isochrones zeroed to a solar color of $(B-V)_o$ = 0.65 at an age of 4.6 Gyrs. The observed isochrone has been adjusted for a reddening of $E(B-V)$ = 0.20 and an apparent modulus of $(m-M)$ = 10.25, equivalent to $(m-M)_o$ = 9.63. Note that given the age of the cluster and the cool temperature range of the stars, specific use of the 200 Myr isochrone as opposed to 100 Myr has no impact on the quality of the fit, which is excellent over the $V$ range of interest. A more extensive discussion by \citet[][in preparation]{DE18} produces an age of 150 $\pm$ 25 Myr for $(m-M)$ = 10.16 $\pm$ 0.10, the slight difference in distance tied to the normalization of the solar isochrones. 

By contrast, the mean parallax for the cluster \citep{GA18b} from DR2 implies a true modulus of $(m-M)_o$ = 9.75; the small difference compared to the isochrone fit is well within the current estimates for potential systematic offsets to the DR2 parallax system zero-point \citep{ZI18, ST18}. While the adoption of this value will shift the isochrone down by only 0.1 mag in Fig. 3, it would lie more effectively along the blue edge of the main sequence rather than running through the mean of the points. Additionally, the offset is easily removed by either increasing the reddening to $E(B-V)$ = 0.22 and/or raising the isochrone metallicity by 0.1 dex \citep{TW09}. As we will demonstrate below, even prior to DR2, there was spectroscopic evidence that M35 has a metallicity closer to solar.

\begin{figure}
\figurenum{3}
\includegraphics[angle=270,width=\columnwidth]{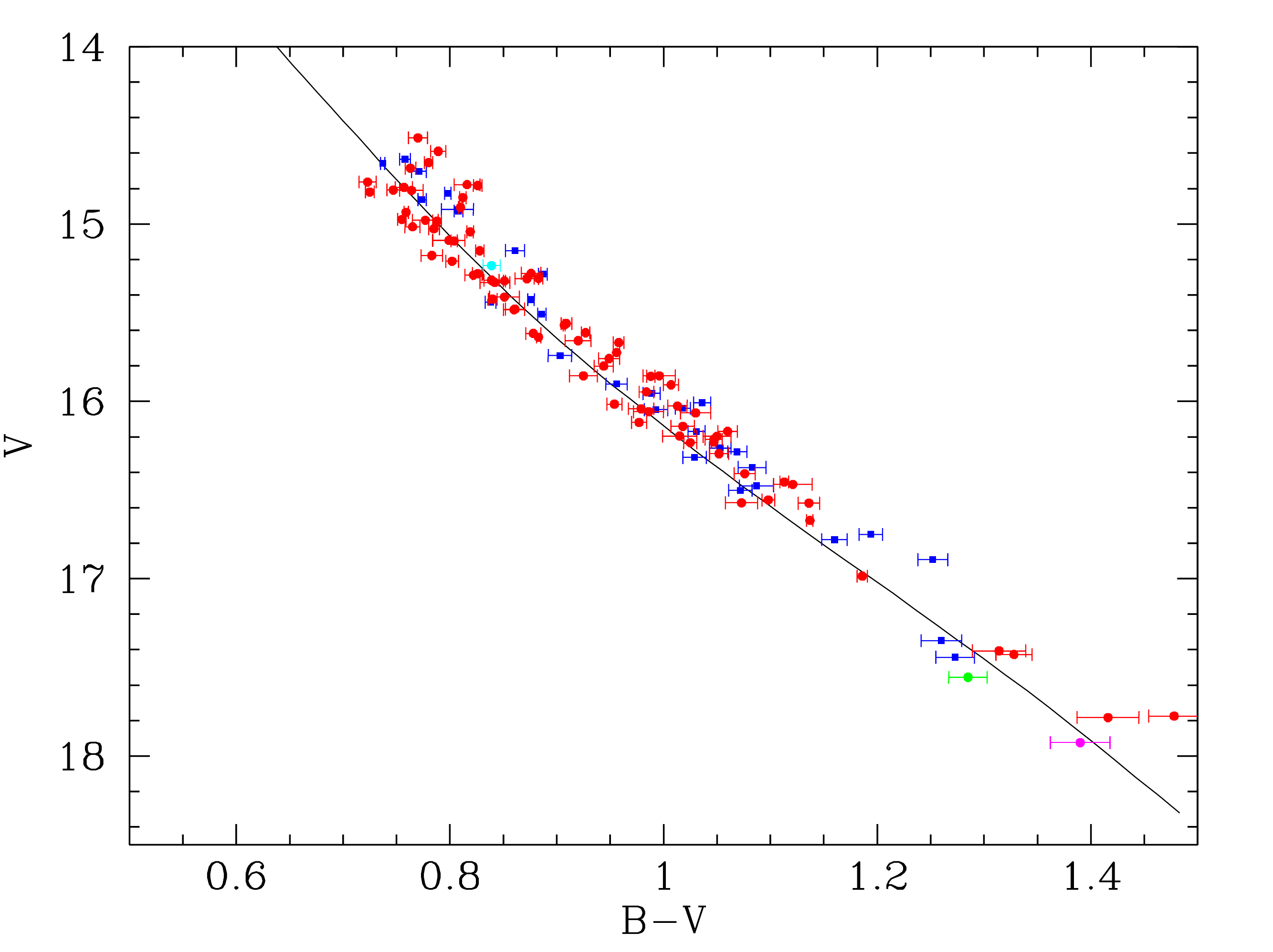}
\caption{CMD for probable members within M35 relative to an isochrone of age 200 Myr with [Fe/H] = -0.14, $E(B-V)$ = 0.20, and $(m-M)$ = 10.25. Symbol colors have the same meaning as in Fig. 2. Filled cyan circle is a binary member, filled green circle is a probable nonmember/binary, and the filled magenta circle has astrometric but not radial-velocity information.}
\end{figure}

\section{Abundances: Li and Otherwise}
{\it Splot}, part of the spectrum analysis suite in IRAF, was used to measure equivalent widths of the line blend 
incorporating the Li-doublet at 6707.8 \AA\ and the Fe I line at 6707.45 \AA; the EW measurements as summarized in Table 3 are an average of three independent measurements of each line, along with the estimated S/N for each spectrum. The restricted wavelength range of the spectra made normal estimation of the S/N per pixel using a line-free region very difficult.  For a few stars, it was possible to estimate the S/N from the spectra; for the remainder, S/N ratios were estimated by extrapolation, scaling by the observed flux levels in the spectra.

\floattable
\begin{deluxetable}{rrrrrrr}
\tablenum{3}
\tablecaption{Measured and Derived Lithium data}
\tablewidth{0pt}
\tablewidth{0pc}
\tablehead{
\colhead{WOCS ID} & \colhead{$B-V$} & \colhead{E.W.} &
\colhead{$S/N$} & \colhead{A(Li)} & \colhead{$\sigma_{ALi}$} &
\colhead{Code} }
\startdata 
8003 & 1.01 & 158.5 & 52 & 2.19 & 0.06 & 0 \\
13006 & 0.82 & 51.5 & 101 & 2.20 & 0.06 & 0 \\
15007 & 0.83 & 120.0 & 101 & 2.70 & 0.03 & 0 \\
16007 & 0.81 & 115.6 & 94 & 2.75 & 0.04 & 0 \\
16010 & 0.77 & 106.6 & 117 & 2.84 & 0.03 & 0 \\
18011 & 0.78 & 101.7 & 108 & 2.77 & 0.04 & 0 \\
21010 & 0.82 & 137.7 & 87 & 2.85 & 0.04 & 0 \\
22030 & 0.76 & 122.4 & 107 & 2.98 & 0.03 & 0 \\
23006 & 0.98 & 120.6 & 50 & 2.09 & 0.06 & 0 \\
24012 & 0.79 & 135.2 & 88 & 2.95 & 0.03 & 0 \\
\enddata
\tablecomments{Codes for Li abundances are 0 for 3-$\sigma$ detections, 1 for upper limit values limited by S/N and 2 for upper limits values limited by the parameter range of the curve-of-growth-based grid.Sample lines for Table 3 are presented here to illustrate the content of the table, which will be available in full online.}
\end{deluxetable}

Before discussing the Li results, a more subtle issue should be addressed. As discussed in the previous section, there are a handful of metallicity determinations for M35, typically leading to the conclusion that the cluster is moderately metal-poor compared to the sun, though just what moderate means ranges from [Fe/H] = -0.1 to -0.3. \citet{BA01} and \citet{ST03b} derived [Fe/H] = -0.21 $\pm$ 0.011 ($\sigma_{\mu}$) and -0.143 $\pm$ 0.014 ($\sigma_{\mu}$), respectively. Our spectra cover a much narrower range in wavelength, minimizing the available Fe line list. As an alternative, however, our spectra do include the CaI line at 6717.68 \AA. This line is consistently stronger (EW $\sim$ 100 - 150 m\AA) than the average Fe line but still overlaps nicely with the strongest line used in the \citet{BA01} analysis, i.e. the line should still lie on or near the linear portion of the curve of growth, especially at solar $T_{\mathrm{eff}}$. As a simple comparison, we have measured the EW of CaI from over 90 solar spectra taken through the same fibers as the M35 stars, generating an average EW of 121.5 $\pm$ 3.4 m\AA\ (s.d.). The same measure has been made for all probable single-star members of M35 as listed in Table 2 by three different measurers and the values averaged. Fig. 4 shows the trend of EW with averaged $(B-V)$. The solid line is the mean relation through the points. The red circle is the solar value, adjusted for a reddening of $E(B-V)$ = 0.20. What is immediately apparent is that the sun sits almost perfectly on the relation defined by the M35 values, despite the fact that, assuming Ca scales as Fe, with [Fe/H] = -0.15, M35 lines should be 30\% weaker than the sun. Any attempt to boost the reddening to 0.25 or higher would shift the sun such that M35 would now be metal-rich compared to the sun. The obvious implication is that M35 may, in fact, be closer to solar metallicity than implied by past measurement. For the discussions of Li which follow, however, we will assume [Fe/H] = -0.15 for the cluster given that the basis for a higher metallicity is tied to a single line.

Our analysis of the Li EW measurements employs a computational scheme that first numerically removes the contribution to the Li line EW produced by the nearby Fe I line at 6707.45 \AA, assuming a cluster metallicity of [Fe/H] = -0.15 and a temperature estimate for the star from the color-temperature relation noted earlier. The computational scheme then interpolates within a model-atmosphere-generated grid of EWs and temperatures to 
estimate a Li abundance for each star, a scheme developed by \citet{ST03b} and employed by \citet{SD04}. A star is only considered to have a detected Li abundance if the corrected EW exceeds three times the estimated error in the EW, itself a function of the measured line width and SNR for the spectrum \citep{DE93, DP93}. 
For stars with EW below this criterion, the Li abundance can only be characterized as an upper limit and no abundance error is estimated. 
The abundance error estimate for detected lines is primarily dependent on the error in the EW. 

\begin{figure}
\figurenum{4}
\plotone{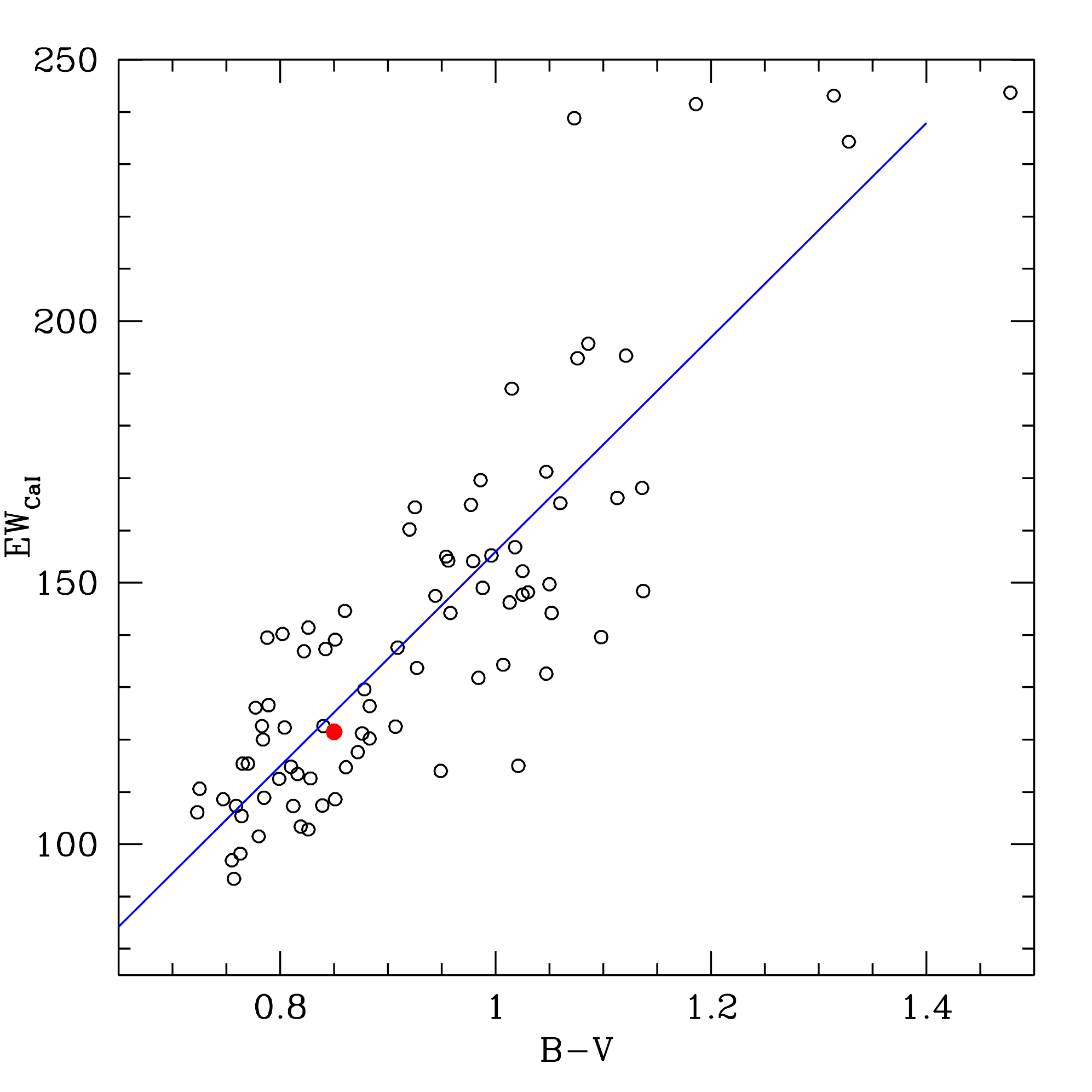}
\caption{Trend of EW measures for the CaI 6717.68 \AA\ line among stars in M35 as a function of averaged, observed pseudo-$(B-V)$ for likely members from Table 2 with measurable A(Li). Blue line is the linear relation through the points. Filled red circle is the sun reddened by $E(B-V)$ = 0.20.}
\end{figure}

Fig. 5 shows the trend of A(Li) as a function of $T_{\mathrm{eff}}$ for the data in Table 3 (red symbols), as well as the EW measures of \citet{BA01} processed through the same procedures as our data (blue symbols). Circles with error bars represent measurable values of A(Li), while triangles without error bars are upper limits. Three points from our investigation have unique designations: a filled cyan circle is the star classed as a binary member, the green square is F8, the potential nonmember/binary, and the filled magenta triangle is the upper limit for the faintest star in the sample with indeterminate membership. If we had adopted a solar metallicity, the mean A(Li) would have been larger by 0.06 dex, though the size of shift is $T_{\mathrm{eff}}$-dependent. At the hot end, A(Li) increases by 0.04 dex but this offset grows to 0.13 dex for the coolest dwarfs with measured A(Li).

\begin{figure}
\figurenum{5}
\plotone{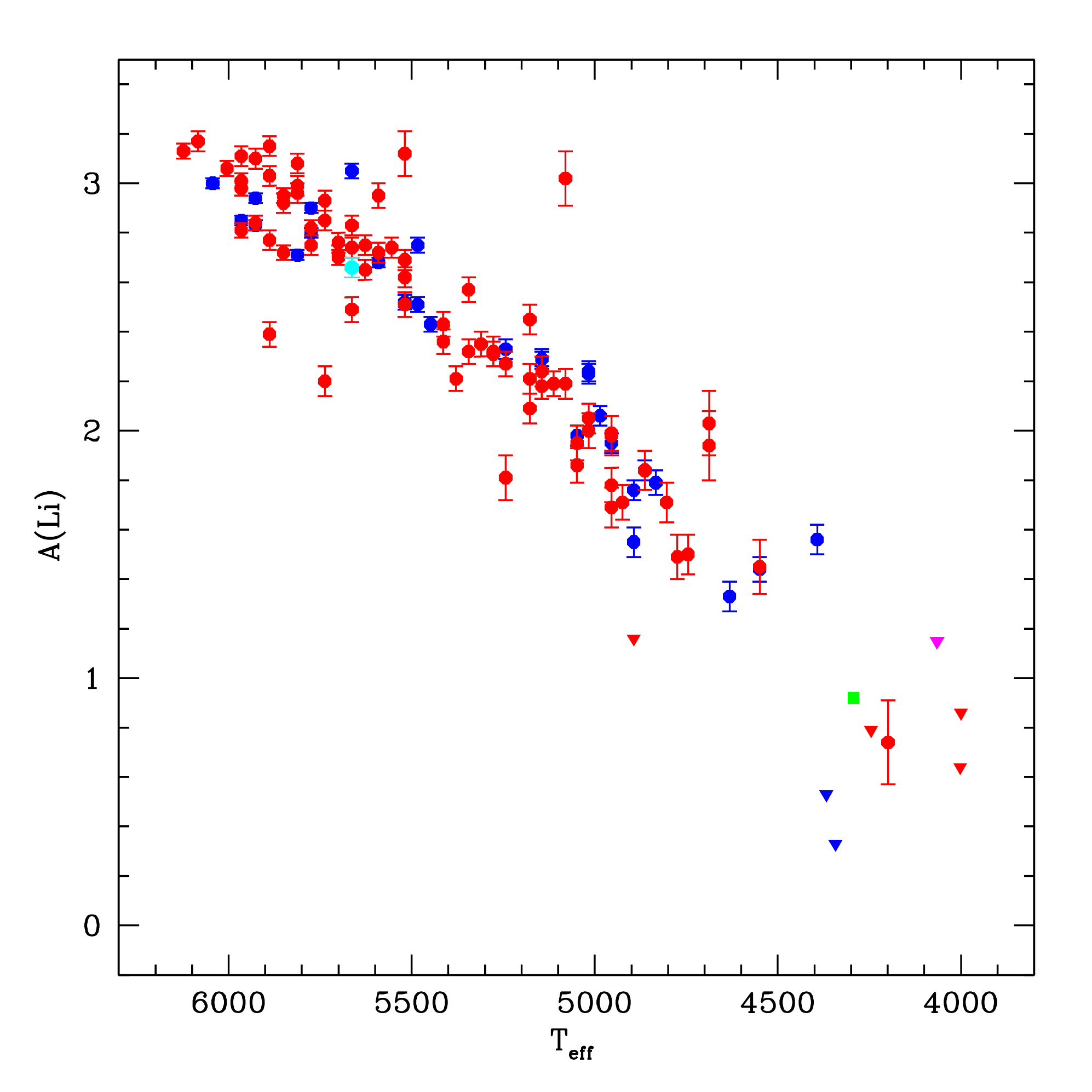}
\caption{A(Li) for likely members from Table 3 (red symbols) and \Citet{BA01} (blue symbols). Triangles are upper limits. The filled cyan symbol is a member binary, the green square is a probable nonmember/binary, and the magenta triangle is the faintest star for which membership is indeterminate.}
\end{figure}

There are a number of key points which should be noted regarding the A(Li) trend. First, the agreement between the patterns defined by each of the two studies, i.e. the overlap among the red and blue points, is excellent, implying that there is no significant evidence for a systematic offset in the EW measures from the two sets of spectra. Second, with the tripling of the single-star sample compared to \citet{BA01}, there is more evidence for real scatter beyond the mean relation, both above and below, at a given $T_{\mathrm{eff}}$.  Previous discussion \citep{BA01} of the scatter on the low A(Li)
side generally assumed that these were likely older field stars with radial velocities comparable to the cluster. With an age of $\sim$150 Gyr for M35,
all field stars with any age above 150 Myr would fall on the more evolved side of the A(Li) relation. The advent of precision astrometric data
from DR2 eliminates that explanation for the stars with low A(Li) that remain in Fig. 5  

Second, prior to DR2, the flip side of the previous argument implied that the stars which scatter on the Li-rich side of the mean relation were, if field stars, significantly younger than M35, which seemed unlikely given an age of $\sim$150 Myr for the cluster. Taking the spectroscopic and photometric scatter into account, we identify 7 stars which lie systematically above the mean sample, 5 from the current study and 2 from \citet{BA01}. What, if anything, provides a common link among these stars? Normally, a higher than normal A(Li) would imply a younger age (or lower metallicity) for these stars compared to those in M35 or a mechanism whereby these stars have retained their Li for a greater length of time than predicted by SSET. Since the degree of convective mixing and Li-depletion are often tied to stellar rotation for main sequence dwarfs, we have made use of the measured $V_{ROT}$ for each star as listed in Table 2, first from the data of \citet{LE15} if available and, if not, from the values measured in the current study or in \citet{BA01}. Based upon the 3 stars in our sample classified by \citet{LE15} as rapid rotators (RR), we have arbitrarily defined any star with a measured $V_{ROT}$ above 20.0 km/sec as a probable rapid rotator. Fig. 6 repeats the A(Li) distribution of Fig. 5, using only stars with measured abundances and making no distinction between our data and that of \citet{BA01}. Stars classed as probable rapid rotators are filled magenta symbols while all others are dark blue. Only 8 stars meet our criterion for rapid rotation; of the 7 stars with high Li, 4 fall within the rapid rotator class. Since $V_{ROT}$ includes a {\it sin i} term, the absence of rapid rotation for 3 of the stars does not mean that they are rotating slowly.

\begin{figure}
\figurenum{6}
\plotone{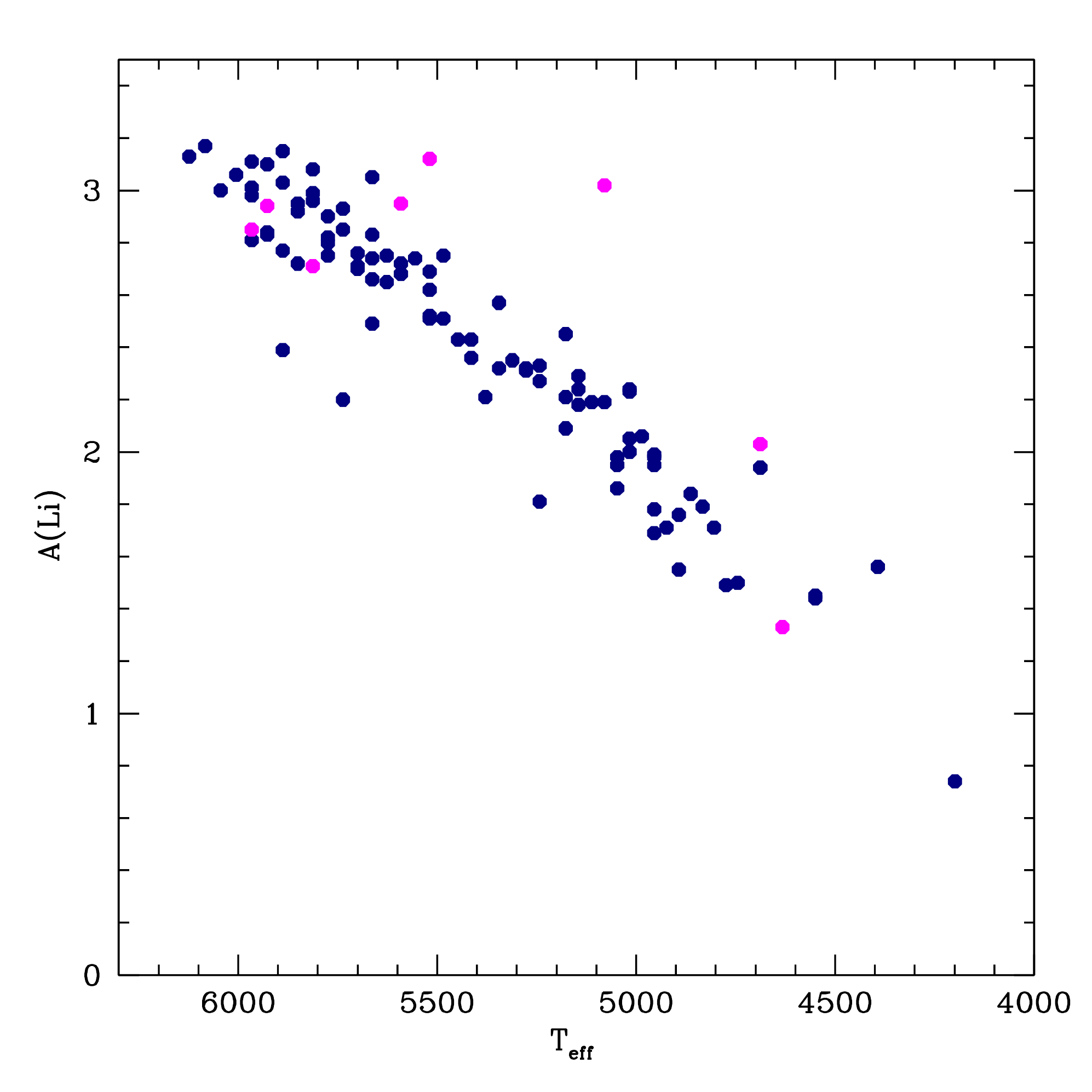}
\caption{A(Li) for likely members from Table 3 and \Citet{BA01}. Only stars with measured A(Li) are included. Dark blue symbols are stars with $V_{ROT}$ below 20.0 km/sec while filled magenta points are stars classed as rapid rotators.}
\end{figure}

As an alternative source of rotation information and, indirectly, insight into Li evolution within the cluster, use can be made of the extensive photometric variability survey of M35 by \citet{ME09}. From a sample of 441 stars with photometrically-derived rotation periods, \citet{ME09} isolate 310 likely members of M35; 16 single-star members of \citet{BA01} and 36 single-star members from Table 3 are included in the survey. Eliminating one star with only an upper limit to A(Li), we plot in Fig. 7 the A(Li) distribution as a function of mean $(B-V)_o$ for the \citet{BA01} data (blue symbols) and Table 3 (red symbols). The filled symbols are stars classed as C (convective) while open symbols are stars classed as I (interface) \citep{BA03}. The classification is critical in that C stars are those which have retained unusually short rotation periods at a given color, i.e. they are rapid rotators. The I stars follow a diagonal distribution with increasing period at redder $(B-V)_o$, indicating that these stars have spun down from an initial value which placed them originally within the C band. As the sample ages, the fraction of stars populating the I band rises and the typical period for a star within the I band at a given color grows longer. Stars caught between these two relations are referred to as G (gap) stars. It is this changing distribution which allows \citet{ME09} to place M35 at an age near 150 Myr, intermediate to that of the Pleiades ($\sim$ 100 Myr) and M34 ($\sim$ 200 Myr). 

The filled circles in Fig. 7 confirm rapid rotation for three of the four red circles sitting above the mean relation in Fig. 6. Star 62041 wasn't included in the \citet{ME09} sample but has a {\it Rapid Rotator} designation in \citet{LE15} and would likely have been classed as a C star, like the other 2 RR stars in Fig. 7. An additional Li-rich star, F4, with $V_{ROT}$ below our limit in Fig. 5, is now included in the C class, overcoming the impact of the $sin$ $i$ term. The one C star (filled symbol) in Fig. 7 located below the mean relation at low A(Li) for its color is WOCS 46015. Consistent with the claimed high rotation speed, 46015 is the only source in our Li sample tagged as an X-ray source by \citet{LE15} based upon the work of \citet{GO13}, who finds that 9 of the 10 single-star X-ray members of M35 are located within the C class. The X-ray emission is normally attributed to unusual levels of chromospheric activity. Whether the low A(Li) is a byproduct of the unusual chromospheric activity is unknown.

Another facet of Fig. 7 which deserves comment is the bimodal distribution of stars into C class or I class. \citet{ME09} find a distribution among the stars in M35 closer to 26\%, 21\%, and 53\% for C, G, and I, respectively. The absence of any gap (G) stars and the deficiency of C stars in Fig. 7 is likely an indication of a selection bias in the original compilation of the sample, potentially tied to the use of photometric criteria in eliminating potential nonmembers. This can be tested by sampling the stars of \citet{ME09} with cluster membership from DR2,  which also overlap with the radial-velocity survey of \citet{LE15} and are classed as single-star members from the latter survey. We find 114 stars, 2 of which failed to have a classification of C, G, or I. Of the remaining 112, 15 (10, 5) were classed as (C, G) and 97 were I. Using the averaged (B-V) values from the multiple color indices, 9 stars fall within the cooler band of stars in Figure 2 that would be considered potential binaries/nonmembers. Of these 9, 4 were classed as C or G, and only 5 were I. In short, the stars with supposedly more rapid rotation rates than the typical star on the main sequence of M35 are five times more likely to be excluded from our sample based upon photometric selection criteria. It is probable that expansion of the Li sample to include a wider array of redder, supposedly single stars would enhance the population of both C and G stars and add significant scatter to Fig. 7 above the mean relation defined by the I stars. 

\begin{figure}
\figurenum{7}
\includegraphics[width=\columnwidth,angle=270,scale=0.80]{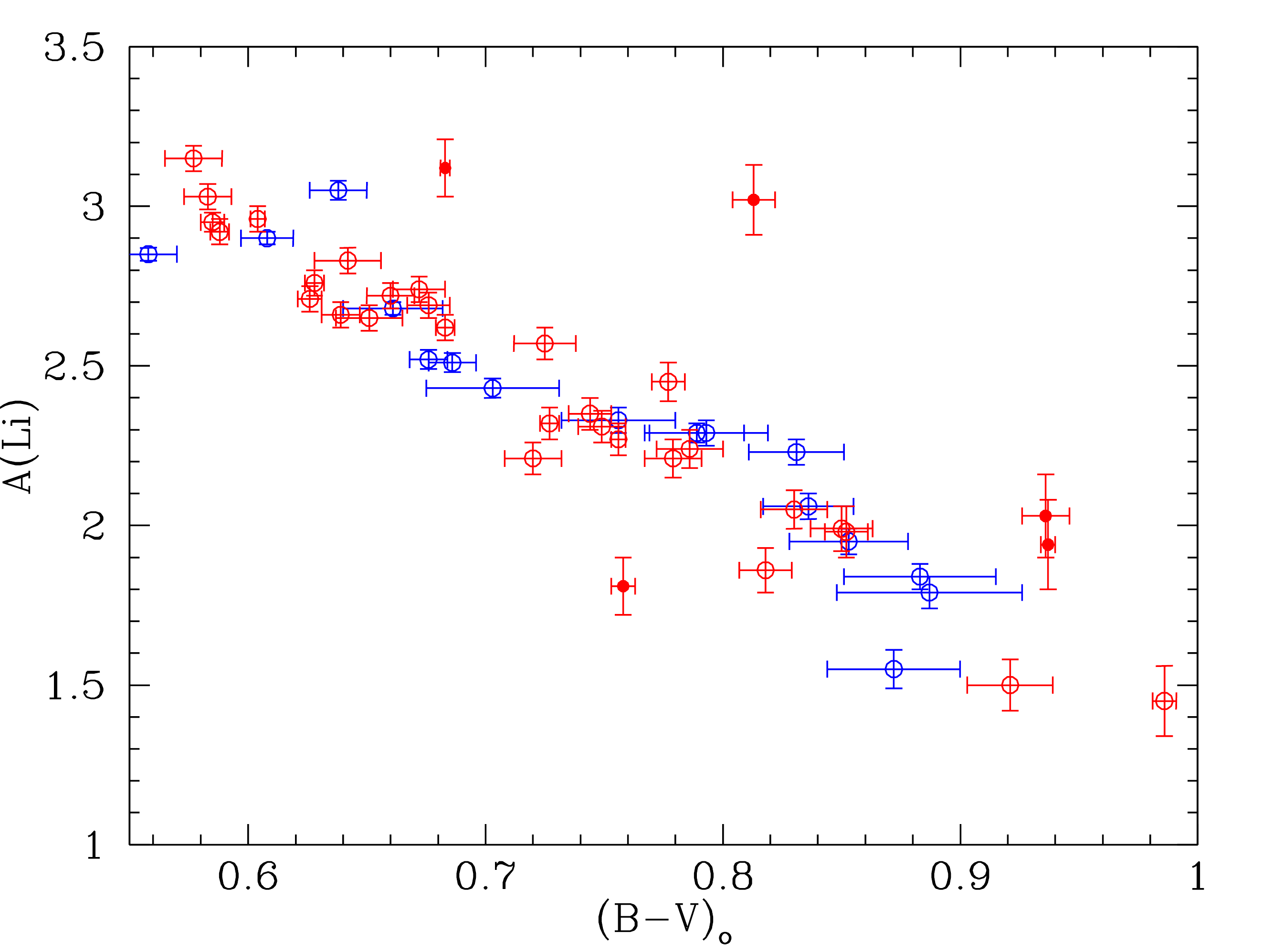}
\caption{A(Li) as a function of intrinsic color for the stars of \citet{BA01} (blue symbols) and Table 3 (red symbols). Open symbols are stars classed as I by \citet{ME09} while filled symbols are classed as C.}
\end{figure}

\subsection{A(Li) with $T_{\mathrm{eff}}$: Comparison to Other Clusters}
While Li measures have been a common theme of cluster studies for decades, we will limit our comparisons of M35 to two primary samples, those in the Hyades/Praesepe database \citep{CU17} and in the Pleiades survey of \citet{BO18}. The choice of these two studies is based upon the ages of the clusters, which should bracket M35, the precision and consistency of the abundance analysis, the identification of probable single and binary members from both spectroscopy and photometry, and the availability of rotation information from either spectroscopy and/or photometry. For the extensive background underlying the previous Li analyses in these clusters, the reader is referred to the two papers noted above.

For the Hyades and Praesepe, the more extensive \citet{CU17} discussion tracks the current analysis closely. The spectra were obtained with Hydra but with a different grating. Precision $UBVRI$ photometry was adopted to identify probable photometric binaries/anomalies and generate $T_{\mathrm{eff}}$ for use in extracting A(Li). The color-$T_{\mathrm{eff}}$ relation adopted was the same applied here. The dominant difference between the two studies is the ability to derive spectroscopic metallicities from the larger bandpass of the spectra, demonstrating that the Hyades and Praesepe have identical [Fe/H] = +0.15, within the uncertainties.

Fig. 8 shows the trend of A(Li) with $T_{\mathrm{eff}}$ for the combined Hyades/Praesepe sample (dark green symbols) and that of M35 (purple symbols). All stars classed as spectroscopic or photometric binaries have been eliminated and only stars with measured A(Li) are included. It is predicted that because of the higher [Fe/H], the Hyades/Praesepe stars likely formed with a higher primordial Li abundance relative to M35. As discussed in \citet{AT18}, the cluster data of \citet{CU11} produce a relation for the primordial cluster A(Li)
 between [Fe/H] = -0.2 and +0.1 with a slope of 0.96. If M35 is more metal-poor by 0.30 (0.15) dex, the dark green symbols should be shifted down by 0.29 (0.14) dex. For purposes of the current discussion, we are more concerned with the trend and dispersion, so no shift in A(Li) has been applied to the Hyades/Praesepe sample. However, with a modest shift in A(Li), all stars in M35, even the pair of Li-deficient stars (purple points) in Fig. 8 which lie on or below the mean trend for Hyades/Praesepe, are encompassed within the range defined by the older cluster pair, placing a comparable upper limit to the age of all the stars within the M35 sample, again assuming all purple points are members.

Beyond the obviously increasing offset between the two datasets with decreasing $T_{\mathrm{eff}}$, expected given the greater age of $\sim$650 Myr \citep{CU17} for the Hyades/Praesepe pair, the other striking difference is the almost total lack of scatter on the high Li side of the distribution for the combined older clusters. While some of the reduced scatter among the mean relation for the Hyades/Praesepe stars is simply due to the greater number of higher precision spectra and the better photometry for the brighter stars observed in these clusters compared with M35, the absence of stars systematically above the mean relation indicates an almost total absence of this class of star within these clusters. Since we have claimed that rapid rotation is linked to the anomalies in M35, what do the Hyades/Praesepe stars tell us? Of the 28 single stars with $V_{ROT}$ above 20 km/sec in the Hyades/Praesepe, none lies at a $T_{\mathrm{eff}}$ below 6200 K. In fact, \citet{CU17} show that the rapid rotators populate the cool side of the Li-dip, with the typical $V_{ROT}$ increasing with increasing $T_{\mathrm{eff}}$ toward the center of the Li-dip. 

\begin{figure}
\figurenum{8}
\plotone{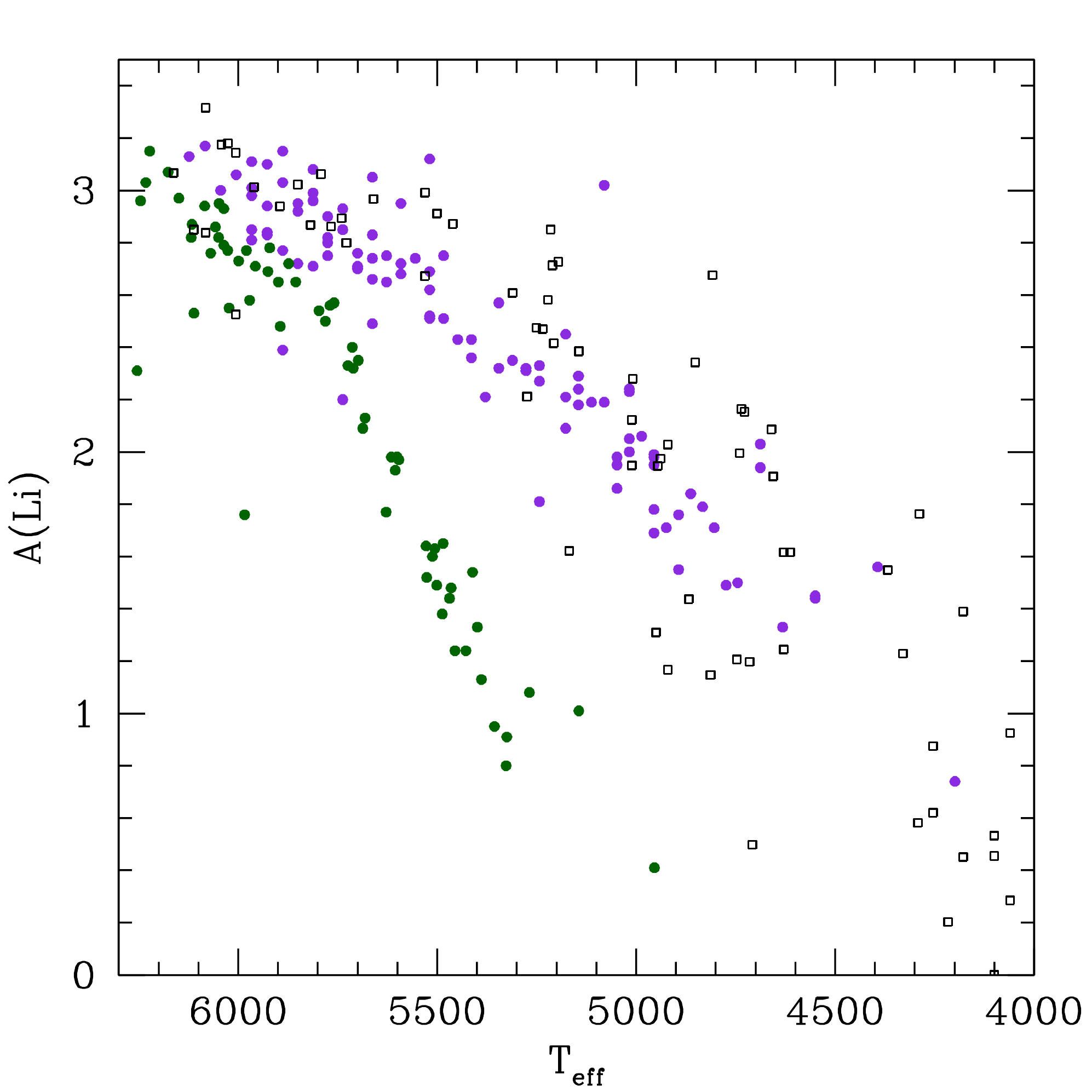}
\caption{A(Li) for likely members of M35 from Table 3 and \citet{BA01} (purple symbols). Dark green symbols are single-star members of the Hyades/Praesepe sample from \citet{CU17} while open black squares are likely single-star members of the Pleiades \citep{BO18}.}
\end{figure}

Moving to a cluster closer in age to M35, the open black squares of Fig. 8 show the single-star members of the Pleiades (age $\sim$ 100 Myr), having removed all binaries, spectroscopic, visual, or photometric \citep{BO18}. The current definitive metallicity for the cluster is [Fe/H] = +0.03, based upon spectroscopic analysis of 20 slow rotators with temperatures near that of the sun \citep{SO09}. Thus, the primordial A(Li) for the Pleiades should have been between 0.17 and 0.03 dex higher than that of M35, assuming the latter cluster had [Fe/H] = -0.15 or 0.0, respectively. As with the Hyades/Praesepe sample, no shift has been applied in Fig. 8. Of the 125 stars in the sample cooler than $T_{\mathrm{eff}}$ = 6200 K, 46 are eliminated as binaries. Of these, 34 would have been removed if only photometric criteria had been available to tag potential anomalous stars. 

Comparing the two relations, M35 and the Pleiades overlap nicely between 6000 K and 5700 K. Between 5700 K and 5100 K, the Pleiades sample falls within the same range of A(Li) as the full spread of stars in M35, with the Pleiades with the lowest A(Li) at a given $T_{\mathrm{eff}}$ 
overlapping with the better defined relation in M35. As expected for a cluster of younger age, however, the typical Pleiades star lies above the mean 
relation for M35. The biggest discrepancy between the two clusters kicks in below $T_{\mathrm{eff}}$ = 5000 K. Again, the limiting range of points at the high A(Li) boundary appears to be the same. The excessive scatter in the Pleiades arises from the low A(Li) points, i.e. stars which even without a metallicity adjustment fall below the mean trend for M35. Identified within a field star sample, these stars would be classed as older than M35, but 
younger than the Hyades, comparable in fact to stars in M34 \citep{GO14}. Between $T_{\mathrm{eff}}$ = 5000 K and 4500 K, there are 8 stars with A(Li) below 1.5. From \citet{BO18}, all 8 have rotation periods above 7.5 days, with an average of 8.04 $\pm$ 0.37 (s.d.) days and a mean A(Li) of 1.15 $\pm$ 0.28 (s.d.) at an average $T_{\mathrm{eff}}$ of 4794 K $\pm$ 113 K (s.d.). All 8 stars would be found within the I class for the Pleiades. The remaining 12 stars in this temperature range have a mean A(Li) = 2.04 $\pm$ 0.29 (s.d.), with periods ranging from 0.5 to 8.5 days, i.e. they would be a mix of C, G, and I stars. 

How does this compare with M35? From the full sample, in the same $T_{\mathrm{eff}}$ range, only the 2 coolest stars out of 17 in the bin fall below 
A(Li) = 1.5, as part of the natural decline in the mean relation with increasing color. If we select only stars which have rotational periods 
from \citet{ME09}, 11 remain. Of these, two were classified as C class members and both have above average A(Li) for their temperature; the remaining 9 are 
all I class and define the mean trend in M35 at these temperatures. The average A(Li), period, and $T_{\mathrm{eff}}$ for these 9 stars 
are 1.79 $\pm$ 0.23 (s.d.), 6.48 $\pm$ 0.60 (s.d.) days, and 4859 K $\pm$ 139 K (s.d.), respectively. The longest period among the 9 stars is 7.52 days.

One possible solution to the scatter of Pleiades below the mean relation for M35 is that the M35 stars are too cool, i.e. the purple points in Fig. 8 should 
be shifted to higher $T_{\mathrm{eff}}$ so that the lower bound in A(Li) at a given $T_{\mathrm{eff}}$ is the same for both clusters and all the scatter shifts to data above the mean relation. This would leave the agreement between the clusters at solar $T_{\mathrm{eff}}$ unchanged since the trend flattens at
these temperatures but enhance the differential for points systematically above the mean M35 trend at cooler $T_{\mathrm{eff}}$. Such a shift would occur if a reddening of $E(B-V)$ = 0.25 had been adopted, making the typical M35 star $\sim$150 K hotter. Unfortunately, raising $T_{\mathrm{eff}}$ by this amount also leads to a re-evaluation of the metallicity (making M35 more metal-rich than the sun) and the Li abundance, collectively raising A(Li) by just over 0.3 dex in the temperature range of interest and leaving the same Pleiades scattered below the M35 relation. 

Rather than looking for a solution to the discrepancy between M35 and the Pleiades within M35, it might be simpler to assign its origin to the Pleiades.  It has been known for decades that the Pleiades dwarfs, when compared to the older Praesepe/Hyades clusters, exhibit a relative position in the CMD which is either too blue, normal, or too red at a given $M_V$, depending on the color index used to define the temperature.  \citet{ST03} proposed that these apparently contradictory results are the byproduct of a spectral energy distribution dominated by spots covering a significant fraction of the stellar surface, often driven by rapid rotation. Thus, the $(B-V)$ index for the Pleiades is not measuring the same surface parameter as the composite, multicolor $(B-V)$ for M35, placing the Pleiades stars in a position which appears too hot/blue for their A(Li).  If coupled with a higher Li-depletion rate for stars with higher [Fe/H], this could significantly impact the scatter defined by the Pleiades stars, at least in comparision to M35. We will return to this point, as well as the potential role of radius inflation, in Section 4.2.

We close this section be recalling the pair of stars in M35 which lie significantly below the mean A(Li) relation as discussed in the previous section. Given their
membership confirmation by DR2, despite the fact that they are typically hotter than the stars defining the large scatter found below 5000 K, this small sample could represent the M35 analog to the Pleiades stars which have spun down to a significantly lower A(Li) for their $T_{\mathrm{eff}}$.

\subsection{A(Li) With Rotation: Comparison to Theory}
Observationally, the G and K dwarfs of M35 appear to be more similar to the Pleiades than previously thought, exhibiting a less populated version of the correlation found between rapid rotation and high A(Li) seen in the younger, latter object. Until recently, this confounding correlation has eluded explanation, not to mention agreement over its existence,  since its discovery over 30 years ago \citep{BU87, SO93}. It now seems possible that the high A(Li) are related to radius inflation that is correlated with rapid rotation; the arguments go as follows.

It has been known for a long time that high stellar activity is correlated with rapid rotation and, in fact, it is highly probable that rapid rotation {\it results} in high magnetic activity.  Models predict that magnetic activity may increase the radii of cool dwarfs either by inhibiting convection \citep{MU01, FE14} or through the effects of dark, magnetic starspots in blocking the emergence of radiative flux at the photosphere \citep{SP86, MA13, JA14}.  Indeed, \citet{JA18} have found that rapidly rotating K and M dwarf Pleiades have average radii that are 14 $\pm$ 2\% larger than the non-rotating models of \citet{DO08} and of \citet{BA15}, and than the interferometric radii \citep{BO12} of older, magnetically inactive field M-dwarfs.  Neither of the two mechanisms listed above can individually explain this over-radius, but it is possible that a combination of both might. Some evidence for radius inflation exists also for other young clusters \citep[][see the discussion in \cite{JA18}]{JA09, JA14, JA16}.

Radius inflation has been proposed as an explanation for: a) the discrepancies between the measured radii, masses, and luminosities of pre-MS and ZAMS stars, and model predictions \citep{KR15, KR16, DA16}; b) anomalous colors of rapidly rotating pre-MS and ZAMS stars \citep{ST03, KA14, CO16}; and c) the present topic of interest, namely the high A(Li) observed in rapid rotators in the Pleiades and other young open clusters \citep{SO14, SO15a, SO15b}.  It should be noted that rapid rotation complicates derivation of accurate A(Li) because of various rotation-related uncertainties.  Our conversion of EW(Li) to A(Li) depends most sensitively on the $T_{\mathrm{eff}}$ (as opposed to on log $g$ or micro-turbulent velocity) for the chosen model atmospheres, yet the degree to which rapid rotation decreases the $T_{\mathrm{eff}}$ is uncertain.  Furthermore, the structure of non-rotating model atmospheres may differ significantly from real rapidly rotating atmospheres, and there may also be significant differences due to the presumed significant presence of spots.  Nevertheless, in spite of these and other uncertainties, we will assume that A(Li) for rapid rotators are indeed cooler and above the A(Li)-$T_{\mathrm{eff}}$ trend for slow rotators, at least for dwarfs cooler than Sun.  The question then becomes whether models can reproduce the general patterns of the A(Li)-$T_{\mathrm{eff}}$ relations for slow versus for rapid rotators, rather than possible fine details of these patterns, which are uncertain.

\citet{SO15b} considered a wide variety of rotation/inflation prescriptions including, for example, a variety of disc-locking times, and reached a number of interesting conclusions: rapid rotation alone with no inflation {\it increases} Li depletion, in contradiction to the Pleiades data, whereas inflation alone, modelled by a reduction in the mixing length parameter, decreases Li depletion.  Three classes of models of combined rotation/inflation effects are contradicted by the data, namely no inflation in all stars, independent of rotation, inflation of 10\% in all stars, independent of rotation, and inflation uncorrelated with rotation. However, models where radius inflation is correlated with rapid rotation reproduce the median Pleiades A(Li)-$T_{\mathrm{eff}}$ pattern, and also the significant under-depletions in A(Li) in stars cooler than the Sun, which may lie as much as 1-2 orders of magnitude above the A(Li)-$T_{\mathrm{eff}}$ trend of slow rotators.  This agreement is insensitive to the precise details of the adopted rotation/inflation prescription.  If these predictions also hold for the older and possibly slightly metal-poor M35, then they may also explain the patterns shown in Figure 6, where most rapidly-rotating stars (magenta dots) lie above the mean A(Li)-$T_{\mathrm{eff}}$ trend of slower rotators (dark blue dots).  Another prediction of these rotation+inflation models is that for stars slightly hotter than the Sun (5770 - 6100 K), the rapid rotators should dip {\it below} the A(Li)-$T_{\mathrm{eff}}$ trend defined by slow rotators.  The Pleiades data are unclear on this point, as rapid rotators exist near 6000 K both below and on the A(Li)-$T_{\mathrm{eff}}$ trend of slow rotators, but, remarkably, {\it all three} rapid rotators in M35 with $T_{\mathrm{eff}}$ between 5770 and 6000 K lie slightly below the A(Li)-$T_{\mathrm{eff}}$ trend of slow rotators, consistent with the models.

Given the apparent agreement between this type of model and another cluster, theoretical exploration of what might happen at older ages could prove informative. In this context, the rapid rotators in the Pleiades (and M35) are explained partly by less actual destruction of Li, which might naively suggest that such stars will always have higher A(Li) than the A(Li)-$T_{\mathrm{eff}}$ trend of slower stars.  However, eventually these stars will presumably spin down and lose more angular momentum than the slow rotators do \citep{PI90}, perhaps thus mixing more and depleting {\it more} Li than slower rotators do.  Which effect ultimately wins, and what is the timing?  Do we expect such stars to lie above or below the A(Li)-$T_{\mathrm{eff}}$ trends of older clusters? At minimum, an expanded sample of cooler stars with A(Li) estimates within M35 could go a long way toward illuminating the appropriate models to resolve these questions.

\section{Summary and Conclusions}

Within the study of Li evolution among open clusters, M35 has a long history, if nothing more than as a counterpoint to the richly diverse A(Li) - $T_{\mathrm{eff}}$ distribution found among the Pleiades. A partial list of the challenges to the claim of a wide intrinsic range of Li \citep{BU87, SO93} among cool Pleiades dwarfs includes chromospheric activity \citep{HO95}, spectroscopic line anomalies \citep{RU96}, magnetic activity \citep{JE99}, atmospheric effects \citep{KI00}, stellar activity and variable reddening \citep{XI05}, and  magnetic fields \citep{LE07}, to name a few. As additional investigations have rendered the claims either inconclusive, inaccurate, and/or inadequate to explain the dispersion, more emphasis has been placed on probing the actual mechanism for Li-depletion among cooler dwarfs, depletion which clearly exceeds the predictions of SSET \citep[see, e.g.][]{XI09, KI10, MI12, GO14, SO15a, SO17}. Thanks to the comprehensive overview supplied by \citet{BO18}, there now appears to be little doubt that the dispersion in Li among the cooler stars in the Pleiades is real and that it is strongly correlated with the cumulative rotational history of the individual cluster stars, with higher A(Li) retained longer by stars with higher $V_{ROT}$, whatever the mechanism by which the likelihood of a spindown is reduced.

The recent discovery of radius inflation in rapidly-rotating Pleiades \citep{JA18} lends strong support to the models of \citet{SO15b} that consider radius inflation correlated with rapid rotation.  Models with rapid rotation alone destroy too much Li, and those with inflation alone decrease Li depletion.  The Pleiades data contradict models with no inflation for all stars, models with 10\% inflation for all stars, and inflation uncorrelated with rotation.  However, models with inflation that correlates with rapid rotation reproduce the mean A(Li)-$T_{\mathrm{eff}}$ relation and also the under-depleted A(Li) in rapid rotators, for stars cooler than the Sun.  Our M35 Li data also support these latter models. In addition, these models predict that rapid rotators just hotter than the Sun (5770 K - 6100 K) should be slightly {\it over-}depleted relative to slow rotators and, while the Pleiades data are silent on this issue, all three M35 rapid rotators in this temperature range lie slightly below the main trend, consistent with the models.

With the expanded sample created by the new Hydra observations coupled with past data for the cooler stars, we can conclude that a large part of the discrepancy between the appearance of M35 and the Pleiades in the A(Li) - color diagram was a result of selection bias in the sample. With too small a dataset contaminated by binaries and nonmembers, the probability of populating the A(Li) - color plot with stars of more extreme $V_{ROT}$ and, indirectly, higher A(Li), was severely reduced. Even with our larger sample of single-star members, selected in part by position in the CMD to maximize cluster membership, the fraction of stars which fall into the C and G rotational bins, rather than  the more slowly rotating I class, is below 10\%, whereas the analysis by \citet{ME09} implies a typical cluster sample between 25\% and 45\%. 

However, the modest sample of rapid rotators does populate a range on the high A(Li) side of the mean relation for M35 which is compatible with the scatter found in the Pleiades. It is expected that a more representative survey of M35 stars covering the full range of the rotation plot delineated by \citet{ME09} would fill in many of the gaps seen in Fig. 7 and improve the similarities with the Pleiades. For statistical reasons and due to a smaller age differential with M35, perhaps the best comparison with Fig. 7 is Fig. 3 of \citet{GO14} showing the A(Li) - $T_{\mathrm{eff}}$ distribution for a modest sample of stars in M34. Here again, the sample is composed almost exclusively of stars either in the C or I class, and the C class stars all lie systematically above the I class stars by almost 1 dex in A(Li).

Somewhat surprisingly, the one region where the Pleiades do show greater scatter than M35 is on the low A(Li) side of the mean M35 relation. While it isn't surprising that these stars have lower A(Li) at a given $T_{\mathrm{eff}}$ since they clearly spin at a slower rate than stars at the same $T_{\mathrm{eff}}$ within M35, it remains a mystery why these stars should attain a slower rotation at a younger age than M35. The nagging possibility remains that this is once again a selection effect, i.e. a more comprehensive  survey covering a wider range of I class stars chosen without respect to position in the CMD would identify some members of this rotationally-challenged class. For now, however, the question remains open.

\acknowledgments

The discussion of M35 was enhanced by access to the data for the Pleiades \citep{BO18} supplied by the authors prior to publication online. Partial support for the cluster Li program was provided to BJAT, DLB and BAT through NSF grant AST-1211621 and to CPD through AST-1211699. The revised text made extensive use of the DR2 database, strengthening the conclusions attained in the original manuscript submitted prior to the DR2 release. The referee's thoughtful suggestions for improving the text are gratefully acknowledged. 

\facility{WIYN:3.5m}
\software{IRAF}

\end{document}